\documentclass[a4paper,10pt,twocolumn]{revtex4}
\usepackage{graphicx}
\usepackage{dcolumn}
\usepackage{bm}

\begin{document}

\title{A nearly relaxation-free opto-electronic memory from ultra-thin graphene-MoS$_2$ binary hybrids}

\author{Kallol\ Roy,$^1$\footnotemark[3] Medini\ Padmanabhan,$^1$ Srijit\ Goswami,$^1$\footnote{present address: Kavli Institute of Nanoscience, Delft University of Technology, P.O. Box 5046, 2600 GA Delft, The Netherlands.} T.\ Phanindra Sai,$^1$ Gopalakrishnan Ramalingam,$^2$\footnote{present address: Department of Materials Science and Engineering, University of Virginia, Charlottesville 22904, United States.}
 Srinivasan Raghavan,$^2$ and Arindam\ Ghosh$^1$\footnote[3]{correspondence information: kallol@physics.iisc.ernet.in, arindam@physics.iisc.ernet.in}}

\address{$^1$Department of Physics, Indian Institute of Science, Bangalore 560012, India}
\address{$^2$Materials Research Center, Indian Institute of Science, Bangalore 560012, India}



\maketitle

{\bf Ultra-thin planar heterostructures of graphene and other two-dimensional crystals have recently attracted much interest. Very high carrier mobility in a graphene-on-boron nitride assembly is now well-established~\cite{deanNatnano2010,zomerAPL2011}, but it has been anticipated that appropriately designed hybrids could perform other tasks as well~\cite{novoselovNATURE2012}. A heterostructure of graphene and molybdenum disulphide (MoS$_2$) is expected to be sensitive to photo illumination due to the optical bandgap in MoS$_2$~\cite{heinzPRL2010}. Despite significant advances in device architectures with both graphene~\cite{xiaNatNano2009,xiaNL2009,ghoshAPL2010,dassarmaRMP2011} and MoS$_2$~\cite{radisavljevicNatnano2011,cnrraoACSNANO2012,zongyouACSnano2011,leeNL2012,branimirACSnano2011}, binary graphene-MoS$_2$ hybrids have not been realized so far, and the promising opto-electronic properties of such structures remain elusive. Here we demonstrate experimentally that graphene-on-MoS$_2$ binary heterostructures display an unexpected and remarkable persistent photoconductivity under illumination of white light. The photoconductivity can not only be tuned independently with both light intensity and back gate voltage, but in response to a suitable combination of light and gate voltage pulses the device functions as a re-writable optoelectronic switch or memory. The persistent, or `ON', state shows virtually no relaxation or decay within the the experimental time scales for low and moderate photoexcitation intensity, indicating a near-perfect charge retention. A microscopic model associates the persistence with strong localization of carriers in MoS$_2$. These effects are also observable at room temperature, and with chemical vapour deposited graphene, and hence are naturally scalable for large area applications.}

Many optical switches and memories rely on persistent photoconductivity (PPC) where electromagnetic excitation shifts the electrical resistance of the host material to a slowly relaxing state. Traditionally, PPC is observed in compound semiconductors, alloys and heterojunctions~\cite{kastalskySSC1984,nathanSSE1986,queisserPRB1986}, but more recently, assemblies of carbon-based nano-materials~\cite{julienADMA2006} have also received considerable interest. External tunability of the PPC, for example with a gate, makes such designs particularly attractive for switching and read-write memory applications, as demonstrated with nanotubes dispersed in photoactive polymers~\cite{julienADMA2006}.

PPC has so far not been realized in graphene-based field-effect devices, although  such devices would readily benefit from strong coupling to light over a wide band of wavelengths in graphene~\cite{dawlatyAPL2008,nairSci2008}, and fast recombination life times~\cite{georgeNL2008,urichNL2011,songNL2011,gaborSci2011}, along with gate-tunability, miniaturization and ability to pattern over a large area~\cite{ghoshAPL2010}. A significant improvement in photoconductivity of graphene has been achieved by combining graphene with other light absorbing materials such as quantum dots~\cite{konstantatosNatnano2012} and chromophores~\cite{myungwoongNL2012}. This indicates that exotic optoelectronic response could be engineered in graphene-based hybrids with an appropriate choice of complementary materials.

Molybdenum disulphide (MoS$_2$) is intrinsically responsive to light, owing to its bandgap which increases from about $1.2$~eV (indirect) in bulk/multilayer MoS$_2$ to $1.9$~eV (direct) for a single molecular layer~\cite{heinzPRL2010}. MoS$_2$ can also be exfoliated mechanically, and can be stabilized in single or multilayered form on insulating substrate~\cite{radisavljevicNatnano2011,ghatakACSnano2011}. This opens up an avenue to create planar heterojunctions of Graphene and MoS$_2$ through physical proximity. In this work we have explored the opto-electronic properties of such planar graphene-MoS$_2$ hybrids.

\begin{figure*}
\includegraphics[scale=1]{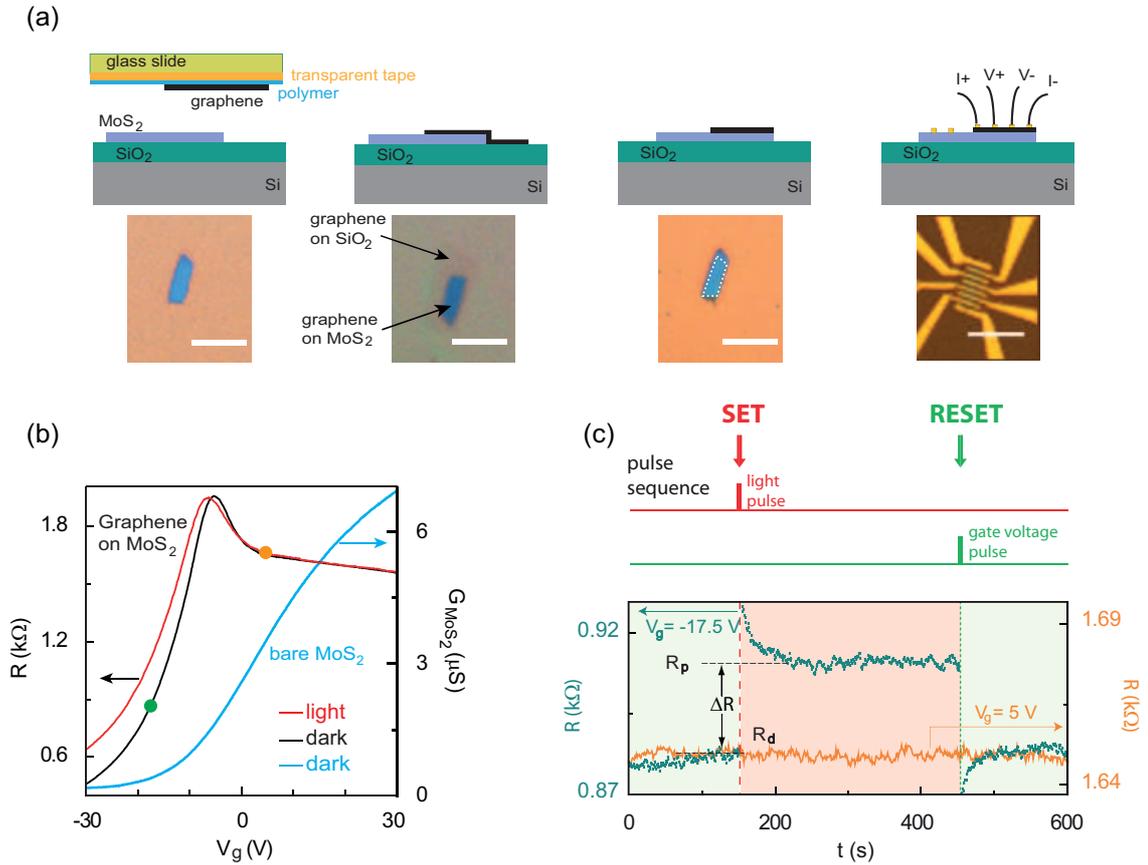}
\caption{\textbf{Device fabrication and basic optoelectronic response.} (a) Schematic of transfer process with corresponding optical images (scale bars are of 10~$\mu$m length). The doted line in the third optical image indicates the outline of graphene. (b) Black and red solid curves represents $R-V_{g}$ traces for graphene on MoS$_2$ in the absence and presence of white light respectively. Cyan coloured trace shows $V_g$ dependent conductance of bare MoS$_2$ measured on a different device in dark. (c) Toggling of the system between two states. $R$ $vs$ $t$ plot is shown illustrating the `set' and `reset' operations achieved by light and gate pulses, respectively. For both curves, $I_{LED}$ = 5 mA. Note that the transient spikes in $R$ following the gate pulses have been removed throughout the manuscript. The vertical red (dashed) and green (dotted) lines are used to indicate the light and gate pulses, respectively.}
\end{figure*}

The device architecture, process flow and measurement layout are shown in Fig.~1a. We have used single layer graphene and multilayer ($\approx 10$ molecular layers) MoS$_2$, both of which were obtained via mechanical exfoliation, and subsequently overlayed by vertical alignment (see methods for details). This technique is same as that for overlaying exfoliated graphene on BN microflakes, which provides a trap-free high-quality interface~\cite{zomerAPL2011,deanNatnano2010}, essential for the present experiments. In all devices, the MoS$_2$ was  exfoliated on a SiO$_2$/p$^+$-Si substrate, and graphene was then transferred on top of MoS$_2$. The layout of the electrical leads can be designed to measure the four-probe resistance ($R$) of graphene (on MoS$_2$), and two-probe conductance ($G_{MoS_2}$) of the underlying MoS$_2$ (that is not covered by graphene) simultaneously. The devices were loaded on a variable-temperature cryostat along with a commercial white light emitting diode (LED) to provide the photo excitation.

Typical gate voltage ($V_g$) characteristics, recorded at 110 K, of the graphene (on-MoS$_2$) and bare MoS$_2$ regions are shown in Fig.~1b. The $R - V_g$ characteristics of graphene, even in the absence of light (black trace in Fig.~1b), was found to be very different from the usual gate transfer characteristics of graphene on an insulating dielectric such as SiO$_2$. The normal `bell-shaped' characteristics are observed only at large negative $V_g$, whereas for positive $V_g$, $R$  becomes  increasingly insensitive to $V_g$. This asymmetric $R-V_g$ can be readily understood by referreing to the $G_{MoS_2} - V_g$ characteristics of the underlying MoS$_2$ (cyan trace). Since we used natural $n-$type MoS$_2$ for exfoliation, the MoS$_2$ channel starts populating only for $V_g \gtrsim V_T$, where $V_T$ is the threshold voltage for conduction. Typically, $V_T$ lies between $-10$~V to $-20$~V for most multilayer MoS$_2$ channels that we have measured~\cite{ghatakACSnano2011}. Here, when $V_g > V_T$, MoS$_2$ turns `ON' (conducting), and screens the gate voltage applied to graphene. Two points need to be noted here: First, for the device presented here, the Dirac point ($V_D = -8$~V) lies close to $V_T$, which is a coincidence, and the two voltage scales can be very different ({\it e.g.} see Fig.~4a, where $V_D = +14$~V). Second, for $V_g \ll V_T$, where MoS$_2$ behaves as a (poor) dielectric, it does not form a parallel conducting channel.

\begin{figure}
\includegraphics[scale=1]{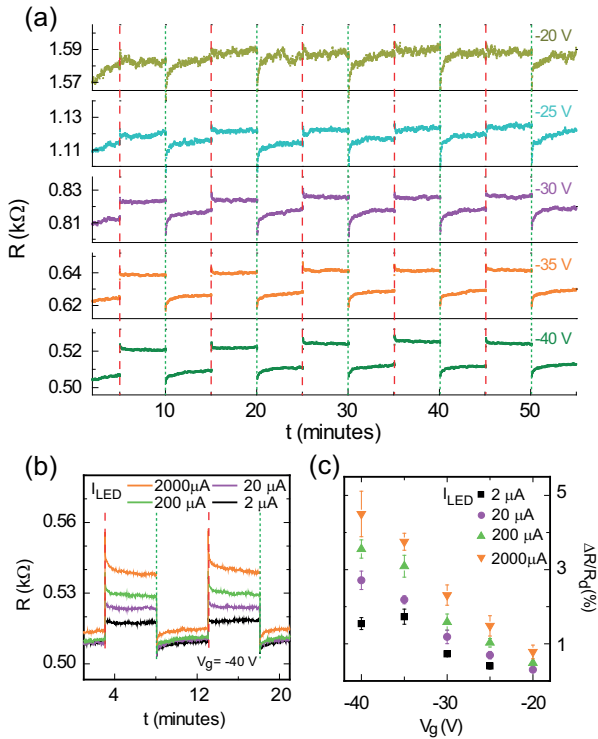}
\caption{\textbf{Gate and light dependent switching.} (a) The evolution of the switching effect as a function of $V_{g}$. For all traces, $I_{LED}$ = 10 $\mu$A. (b) Switching action at different light intensities. (c) Percentage change in resistance as a function of gate voltage calculated  from traces similar to those shown in (b).}
\end{figure}

We now illuminate the device by forward biasing a white LED with a current $I_{LED}$. The red trace in Fig.~1b was recorded while the device was illuminated continuously with $I_{LED}=5$~mA corresponding to a photoexcitation power of $\sim50$~pW/$\mu$m$^2$. Note that there is a pronounced effect only on the negative side of  the $V_T$ where an {\it increase} in $R$ is observed in the presence of light. This negative photoconductivity can be understood as a transfer of the photogenerated electrons inside MoS$_2$ to graphene under the influence of a negative $V_g$. Since graphene is hole-doped in this regime, addition of electrons increases $R$. However, when $V_g \gg V_T$, the bands in MoS$_2$ stay flat due to its near-metallic nature. This suppresses the transfer of photoexcited carriers from MoS$_2$ to graphene.

The response of the hybrid was found to be significantly more striking when we used pulsed, instead of continuous, photoexcitation. Fig.~1c shows the response of $R$ to a sequence of light and gate voltage pulses for two different $V_g$ on opposite sides of $V_T$ (denoted by the markers in Fig.~1b). At $V_g = -17.5$~V (i.e., $V_g < V_T$) the light pulse ($I_{LED}$ pulse height = 5~mA, pulse width~$\sim 0.1$~s) induces a transient-like behavior in $R$. It rises instantaneously  from its dark state value ($R_d$), and then decays over few tens of seconds to a {\it higher} final value ($R_p$), indicating a persistent negative photoconductivity. The magnitude of PPC is defined as $\Delta R=R_p-R_d$. On the other hand, it is clear that when $V_g > V_T$ ($V_g = +5$~V), the light pulse has no measurable effect on $R$, in agreement with experiments performed with continuous illumination. This sensitivity of the PPC to the relative positions of $V_g$ and $V_T$ allowed us to develop a strategy to {\it erase} the PPC in our devices. By introducing a pulse in $V_g$ we were able to `reset' the device to its original dark state resistance from the persistent state. The trace corresponding to $V_g = -17.5$~V in Fig.~1c demonstrates this switching. A gate voltage of pulse height~$\Delta V$ and duration~$\approx~0.5$~s restored $R$ to $R_d$ within a timescale limited by the measurement/pulsing electronics. Typically, we used $\Delta V = +20$~V, so that $V_g + \Delta V \gtrsim V_T$. We would like to point out that our control experiments with MoS$_2$ and graphene separately show no clear evidence of PPC in both cases (see Fig.~S1 in supplementary information (SI)).

Fig.~2a shows the evolution of the PPC cycles with $V_g$ ($I_{LED} = 10~\mu A$). It is clear that $\Delta R/R_d$ is largest for $V_g = -40$~V, and almost vanishes when $V_g$ is increased to $-20$~V. At any $V_g$, the `light set - voltage reset' cycles could be performed over days with better than 95\% accuracy in recovering the PPC magnitude. The PPC magnitude also decreases when the intensity of light pulse is reduced, as can be seen in Fig.~2b, where we varied $I_{LED}$ over three decades of magnitude. The results of Figs.~2a and 2b are summarized in Fig.~2c where $\Delta R/R_d$ is plotted as a function of  $V_{g}$ for various light intensities. $\Delta R/R_d$ is found to be as large as 5\% at T$\sim 110$~K (similar values of PPC can be achieved at room temperature by making $V_g$ more negative). It is perhaps not surprising that the PPC magnitude depends on the light intensity, however its extreme sensitivity to gate voltage opens up the possibility of creating a new class of electrically tunable optoelectronic devices.

A key feature of the switching cycles in Fig.~2 is the absence of a time-dependent decay in the photoconductivity in the persistent state, particularly at low photo-illumination intensity (Fig.~2b). In Fig.~3a we have plotted the photoconductivity in the persistent state in one of the cycles after a low-intensity pulse ($I_{LED} = 2~\mu$A) for three different $V_g$. The PPC shows no decay with time over three decades irrespective of $V_g$, remaining essentially constant even when we monitored it over more than 10 hours (inset of Fig.~3b, here $I_{LED} = 5~\mu$A and $V_g = -40$V have been used). At higher $I_{LED}$ ($\gtrsim 50~\mu$A), we do observe a logarithmic decay at times $< 50$~s, although at long times photoconductivity becomes nearly time-independent again. This is illustrated in Fig.~3b with PPC relaxation for $I_{LED} = 200~\mu$A pulses.

\begin{figure}
\includegraphics[scale=0.9]{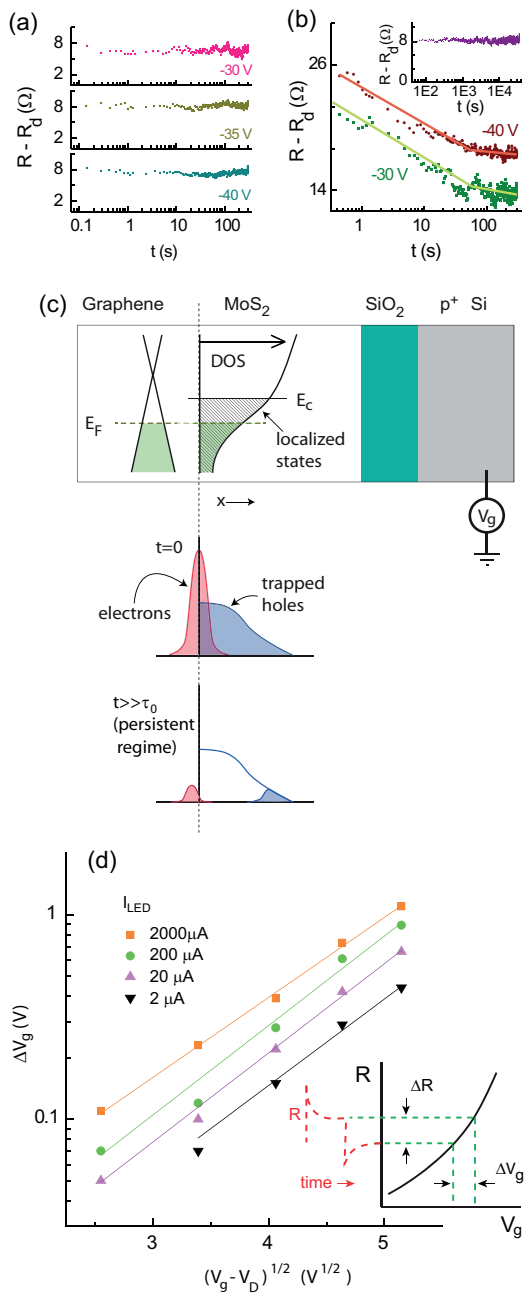}
\caption{\textbf{PPC relaxation nature and mechanism.} (a) Nearly relaxation free nature of persistent states at three different gate voltages. $I_{LED} = 2~\mu$A has been used for 0.1~s to `excite' the system. (b) Transition from logarithmic decay to nearly relaxation free state after photoexcitation with LED current 200~$\mu$A for 0.1s. Solid lines are guide to eye. Inset shows the trace of `ON' state over long time scale ($\sim~12$~hrs). (c) Schematic showing the evolution of electron-hole distribution in the system as a function of time. $t=0$ corresponds to the moment after the light pulse. (d) $\Delta V_g$ $vs.$ $\sqrt{V_g-V_D}$. This plot is calculated using the data shown in Fig.~2c. The inset illustrates the conversion of $\Delta$R to $\Delta V_g$.}
\end{figure}

The near-absence of time decay of the persistent state suggests a strong potential barrier that prohibits recombination of electron and holes created on photo-illumination. To understand this we note that the majority carriers (electrons) in natural MoS$_2$ flakes are strongly localized~\cite{ghatakACSnano2011}, and display Mott-type variable range hopping transport when $V_g$ is reduced to the conduction threshold $V_T$. Thus our graphene-MoS$_2$ hybrid behaves as a heterojunction of a doped conducting system (graphene) and a semiconductor (MoS$_2$) where the carriers are localized, and trapped in potential wells that separate them from the conducting region~\cite{queisserPRB1986}. $V_T$ represents the gate voltage at which the Fermi energy ($E_F$) approaches the mobility edge ($E_c$). The distribution of carriers~\cite{queisserPRB1986} right after the light pulse (t = 0) is shown in the schematic of Fig.~3c. The electrons are swept towards the graphene-Mo$S_2$ interface, and addition of electrons to the p-doped graphene results in an immediate increase in $R$. The holes remain trapped in MoS$_2$, but after light is switched off the hole distribution is modified by: (1) quantum tunneling mediated electron-hole recombination, and (2) thermal excitation of holes from the graphene to MoS$_2$. The PPC then determined by the excess electron density, $n(t)$, left in graphene at time $t$ after the illumination is turned off. As derived in the supplementary information (Eq.~S5), $n(t)$ can be expressed as,

\begin{equation}
\label{Eq1}
n(t)\propto f(t)\exp[(E_C-E_F)/k_BT]
\end{equation}

\noindent where $f(t)$ is a logarithmically decaying function when the initial hole distribution is assumed continuous and uniform. However, $f(t) \rightarrow$ constant for isolated local accumulation of holes far away from the graphene-MoS$_2$ interface, as the recombination time becomes exponentially long. This is consistent with the observation that decay-free PPC occurs only for weak illumination, or at long time when the illumination is strong. Strong localization in MoS$_2$ makes recombination via quantum tunneling or thermal activation virtually ineffective, and the equilibrium is restored only by lifting $E_F$ to $E_C$, which provides the ``erasing'' mechanism of the PPC as $R$ reverts to its dark-state resistance.

To address the $V_g$-dependence, we estimated $n(t)$ from the asymptotic time-independent value of $\Delta R$ and the equivalent shift ($\Delta V_g$) in the gate voltage. The scheme is shown in the inset of Fig.~3d, where $\Delta V_g$ could be evaluated directly from the experimentally measured $R-V_g$ characteristics (black trace in Fig.~1b). Due to the linear bands in graphene (see SI for details), we find:

\begin{equation}
\label{Eq3}
\Delta V_g \propto \exp[\beta\sqrt{|V_g - V_D|}]
\end{equation}

\noindent where $\beta = \hbar v_F\sqrt{(\pi C_{eff}/e)}/k_BT$. Here, $C_{eff}$ is the effective capacitance between the silicon backgate and graphene and $v_F$ is the Fermi velocity in graphene. The exponential increase in $\Delta V_g$  with $|V_g - V_D|^{1/2}$ is clearly observed for different photoexcitations in Fig.~3c. Experimentally, we get $\beta \approx 1.1$ which indicates that $C_{eff}$ is about a factor of $\sim 5$ smaller than the bare capacitance of 285~nm SiO$_2$ that we used as gate dielectric. This can be readily attributed to the quantum capacitance of the MoS$_2$ film which acts as a `leaky' capacitor. Note that, if we consider this modified value of the capacitance, we find that the mobility of graphene in our sample is $\sim 12,000$~cm$^2$V$^{-1}$s$^{-1}$ which is higher than that typically obtained for exfoliated graphene on bare SiO$_2$ substrates.

Finally, we ask three important questions: (i) whether the PPC can be observed in other graphene devices, particularly for large area scalability (ii) can the devices work at room temperature, and (iii) how stable are the devices over long-time operation (suitable for non-volatile memory applications).

\begin{figure}
\includegraphics[scale=1]{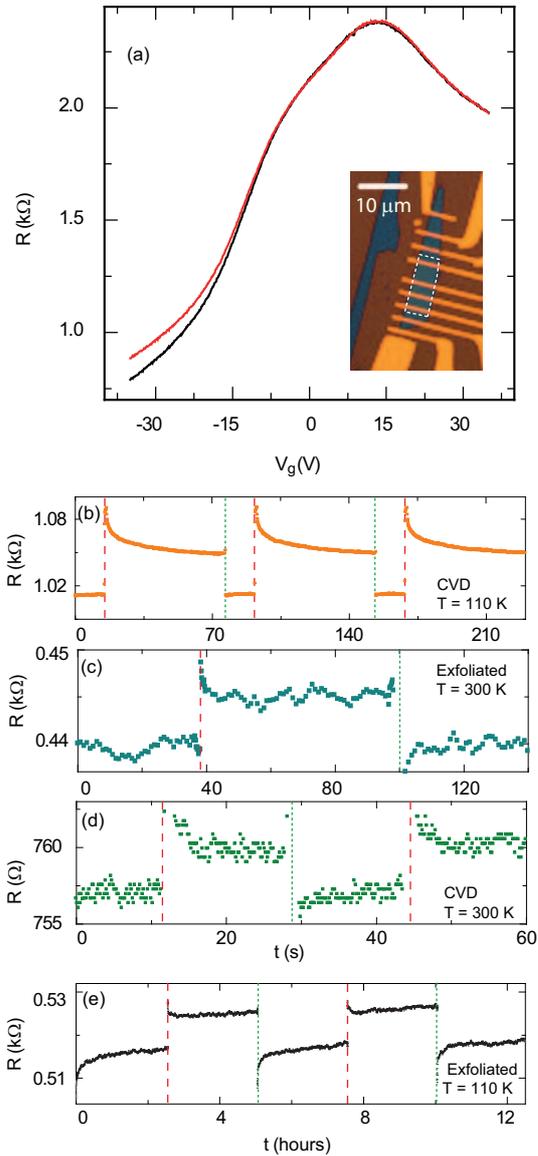}
\caption{\textbf{Scalability, room-temperature operation and long-term retention.} (a) $R-V_{g}$ of a device made of CVD-graphene transferred onto exfoliated MoS$_2$. Inset: optical image of the device where white dashed line is the guide to the graphene outline. (b) Switching action of the device at 110~K. $V_g$ = -30~V and $I_{LED}$ is 5~mA. (c) Data from the exfoliated device at room temperature, $I_{LED}$ = 5~mA. (d) CVD device at room temperature, $I_{LED}$ = 5~mA (e) Data showing the stability of the system in a timescale of hours. Here, $V_g$ = -40~V and $I_{LED}$ = 5~$\mu$A.}
\end{figure}

We have fabricated similar devices by transferring graphene grown via chemical vapor deposition (CVD) onto exfoliated multilayer MoS$_2$. In the inset of Fig.~4a, we show an optical image of one such device. The $R-V_g$ curves of the device with and without light are shown in Fig.~4a. The CVD-device behaves very similarly to the exfoliated device, albeit with a smaller magnitude of $\Delta R/R_d$ for the same intensity of illumination. This might be correlated with the lower mobility and higher disorder typically associated with CVD growth~\cite{matteviJMC2011,atinAPL2010}. Nevertheless, the switching action shown in Fig.~4b confirms PPC. Note that a recent report has shown that large scale CVD growth of MoS$_{2}$ is possible on graphene substrates \cite{shiNL2012}, thereby paving new ways for large area device designs.

In Figs.~4c and 4d, we show that PPC is observable even at room temperature for both exfoliated and CVD-graphene devices. However, thermal broadening reduces the threshold for conduction in MoS$_2$, which makes the effect clearly observable only at large negative $V_g$. For example, the data shown in Fig.~4c corresponds to $V_g = -80$~V. Correspondingly, the gate voltage pulse applied $\Delta V$ ($= +40$~V) for recovery was also large. Fig.~4e shows switching cycles in the exfoliated device over a long period of time extending several hours. It confirms that such devices can be promising for nonvolatile memory applications.

\section{Methods}
\emph{Transfer procedure for laying graphene on MoS$_2$:} Here we follow a similar procedure to Zomer \emph{et al.}~\cite{zomerAPL2011}. A piece of glass is covered with a transparent tape and coated with a polymer such as PMMA or EL-9. Graphene is exfoliated onto this substrate. Separately, MoS$_2$ is transferred to an Si/SiO$_{2}$ substrate by standard mechanical exfoliation and placed on the micromanipulator stage of an MJB3 mask aligner with a custom built heater. The graphene flake, now on a transparent substrate, is positioned upside down and aligned with the MoS$_2$ flake using the microscope of the aligner. The two are brought into contact at an elevated temperature ($T>100^{\circ}$C) as a result of which graphene, along with the polymer is transferred onto MoS$_2$. The polymer is later washed off with acetone.

\emph{Device fabrication and measurement:} Once transferred onto MoS$_2$, graphene (exfoliated or CVD-grown) is etched into a desired shape by oxygen plasma. Contacts are drawn using standard e-beam lithography and metallization is done with Ti/Au (15 nm/45 nm) or Cr/Au (15 nm/45 nm).  In devices where contacts need to be put on MoS$_2$ also, we use Au as the contact metal. Measurements are done on graphene using a standard 4-probe lock-in technique. A commercial white LED is used as the light source.

\emph{CVD growth:} The CVD-graphene reported in this study was synthesized by low-pressure CVD (base pressure of 1 Torr). 25 $\mu$m thick copper foils were annealed at 1000 $^{\circ}$C for 5 minutes under a H$_2$ flow of 50 sccm to reclaim the pure metal surface. CH$_4$ and H$_2$ were then introduced at a rate of 35 sccm and 2 sccm for a growth time of 30 s. The reactor was cooled down to room temperature at a cooling rate of 8$^{\circ}$C/minute under a 1000 sccm flow of H$_2$.

\section{Acknowledgement}
We acknowledge the Department of Science and Technology (DST) for a funded project. S.R. acknowledges support under Grant No. SR/S2/CMP-02/2007.

\end{document}